\nonstopmode
\documentclass[11pt]{article}

\usepackage[fleqn]{amsmath}

\usepackage{hyperref}
\hypersetup{
    colorlinks=true,
    linkcolor=blue,
    filecolor=magenta,      
    urlcolor=blue,
}

\usepackage{tcolorbox}   %% colored background
\definecolor{mycolor}{rgb}{0.122, 0.435, 0.698}% Rule colour

  \usepackage{geometry}
\usepackage{amsmath,amssymb,amsthm}
\usepackage{graphics,epsfig,calc}

\usepackage{latexsym,epsfig,bm,amssymb}
\usepackage{xcolor}
\usepackage{amsthm,mathrsfs}

\DeclareSymbolFont{AMSb}{U}{msb}m{n}
\DeclareSymbolFontAlphabet{\mathbb}{AMSb}

\newcommand{\beqn}{\begin{eqnarray}}
\newcommand{\eeqn}{\end{eqnarray}}
\newcommand{\be}{\begin{equation}}
\newcommand{\ee}{\end{equation}}

\newcommand{\bsp}{\begin{split}}
\newcommand{\esp}{\end{split}}

\newcommand{\ba}{\begin{array}}
\newcommand{\ea}{\end{array}}

\newcommand{\bpr}{\begin{proof}}
\newcommand{\epr}{\end{proof}}

\newcommand{\HS}{{\rm HS}}

\newcommand{\aA}{{\mathbb A}}

\newcommand{\cD}{{\cal D}}

\newcommand{\da}{\dagger}
\newcommand{\ci}{\cite}
\newcommand{\de}{\delta}
\newcommand{{\De}}{{\Delta}}

\newcommand{\fr}{\frac}

\newcommand{\ga}{\gamma}

\newcommand{\la}{\label}

\newcommand{  \om}{  \omega}

\newcommand{ \ov}{ \overline}

\newcommand{\si}{\sigma}

\newcommand\C{{\mathbb C}}
\newcommand\R{{\mathbb R}}

\newcommand{\tr}{\mathop{\rm tr\,}\nolimits}

 \newcommand{\st}{\stackrel}

\newcommand{\toHS}{\st{\HS}{-\!\!-\!\!\!\!\longrightarrow}}

\newtheorem{theorem}{Theorem}[section]

\newtheorem{definition}[theorem]{Definition}

\newtheorem{lemma}[theorem]{Lemma}
\newtheorem{example}[theorem]{Example}

\newtheorem{remark}[theorem]{Remark}

\newtheorem{remarks}[theorem]{Remarks}
\newtheorem{cor}[theorem]{Corollary}
\newtheorem{proposition}[theorem]{Proposition}

\newcommand{\bd}{\begin{definition}}
 \newcommand{\ed}{\end{definition}}

\newcommand{\bt}{\begin{theorem}}
 \newcommand{\et}{\end{theorem}}
\newcommand{\bqt}{\begin{qtheorem}}
 \newcommand{\eqt}{\end{qtheorem}}

\newcommand{\bp}{\begin{proposition}}
 \newcommand{\ep}{\end{proposition}}

\newcommand{\bl}{\begin{lemma}}
 \newcommand{\el}{\end{lemma}}
\newcommand{\bc}{\begin{cor}}
 \newcommand{\ec}{\end{cor}}

\newcommand{\bex}{\begin{example}}
 \newcommand{\eex}{\end{example}}
 
\newcommand{\bexs}{\begin{examples}}
 \newcommand{\eexs}{\end{examples}}

\newcommand{\bexe}{\begin{exercice}}
 \newcommand{\eexe}{\end{exercice}}

\newcommand{\br}{\begin{remark}}
 \newcommand{\er}{\end{remark}}
 
\newcommand{\brs}{\begin{remarks}}
 \newcommand{\ers}{\end{remarks}}

\newcommand{\bce}{\begin{center}}
\newcommand{\ece}{\end{center}}

\begin{document}
%%%%%%%%%%%%%%%%%%%%%%%%%%%%%%%%%%%%%%%%%%%%%%%%%%%
\begin{center}

{\huge On dynamical  semigroup for  damped driven

Jaynes--Cummings equations

}

%\footnote{All data are openly e}
%%AC??
\bigskip\smallskip

 {\large A.I. Komech$^1$ and E.A. Kopylova}\footnote{ 
 Supported partly by Austrian Science Fund (FWF) PAT 3476224.}
 \\
{\it
Institute of Mathematics of
BOKU
University, Vienna, Austria\\
}
 alexander.komech@boku.ac.at,\qquad
 elena.kopylova@boku.ac.at

\smallskip

\end{center}

\setcounter{page}{1}
\thispagestyle{empty}
%%%%%%%%%%%%%%%%%%%%%%%%%%%%%%%%%%%%%%%%%%%%%%%%%%%%%%%%

%\vspace{-15mm}
%\setcounter{page}{1}
%\thispagestyle{empty}
%%%%%%%%%%%%%%%%%%%%%%%%%%%%%%%%%%%%%%%%%%%%%%%%%%%%%%%%%%%%%%%%%%%%%%%%%%%%%%%%%%%%
\begin{abstract}
%%%%%%%%%%%%%%%%%%%%%%%%%%%%%%%%%%%%%%%%%%%%%%%%%%%%%%%%%%%%%%%%%%%%%%%%%%%%%%%%%%%%%

The article addresses the damped driven Jaynes--Cummings 
for quantised one-mode Maxwell field
coupled to a  two-level 
molecule. We consider a broad class of  
 damping and 
 pumping which are polynomial in the creation and  annihilation operators.  
 
 Our
  main result is the construction of a contraction dynamical semigroup  in the Hilbert
space of Hermitian Hilbert--Schmidt operators
  in the case of 
  a nonpositive  dissipation operator and time-independent pumping.
  All trajectories of the semigroup are generalised solutions to the Jaynes--Cummings 
  equations.
  As a key example, we prove nonpositivity 
  for the basic dissipation operator of Quantum Optics.
  
     \end{abstract}
    
  \noindent{\it MSC classification}: 
  81V80,
  %Quantum optics
  	81S05,  	
	%Commutation relations and statistics as related to quantum mechanics (general)
81S08  	
%Canonical quantization
  37K06,  	
  %General theory of infinite-dimensional Hamiltonian and Lagrangian systems, %Hamiltonian and Lagrangian structures, symmetries, conservation laws
  37K40,  	
  %Soliton theory, asymptotic behavior of solutions of infinite-dimensional Hamiltonian systems
37K45,  	
%Stability problems for infinite-dimensional Hamiltonian and Lagrangian systems
  78A40, 
  %Waves and radiation in optics and electromagnetic theory
78A60.
  % Lasers, masers, optical bistability, nonlinear optics
  \smallskip
  

\tableofcontents

\setcounter{equation}{0}
\section{Introduction}
The Jaynes--Cummings equations are one of the  basic models of Quantum Optics,
and it is used for description of various aspects of laser action. 
The survey of the results on the
model without damping and pumping can be found in  
\cite{BJ2007,D1976,LM2021}. 
Various versions of pumping are considered in \cite{AGC1992,BHPSM2024,DKM1994,JA1993,SF2002}.
Damping was introduced for the analysis of quantum spontaneous emission \cite{A1973}--\cite{A1974}, \cite{BR1997,BJ2007,SF2002,VW2006}.
We construct global solutions for all initial values from the space of Hilbert--Schmidt 
operators in the case of time-independent pumping.
\smallskip

Denote 
$X=F\otimes\C^2$, where $F$ is the single-particle Hilbert space
endowed with 
%%%%%%%%%
an orthonormal basis $|n\rangle$, $n=0,1,\dots$,
and  the corresponding annihilation and   creation operators $a$ and $a^\dag$:
\be\la{aa} 
 a|n\rangle=\sqrt{n}|n-1\rangle,\qquad 
 a^\dag|n\rangle=\sqrt{n+1}|n+1\rangle, \qquad [a,a^\da]=1.
\ee
We will consider  a  damped-driven version of  the Jaynes--Cummings equations
 (QRM)
 % \ci[(5.107)]{VW2006}: 
\be\la{JC}
\dot \rho(t)=\aA\rho(t):=-i[H,\rho(t)]+\ga D\rho(t),\qquad t\ge 0,
\ee
where the density operator  $\rho(t)$ of the coupled field-molecule system 
 is  a  Hermitian operator in $X$.
%%%%%%%%
 The Hamiltonian  $H$ is the sum 
\be\la{JCh}
H=H_0+pV,\qquad
\mbox{where}
\quad
H_0:=\om_c a^\da a+\fr12  \om_a\si_3,
\quad
V= (a+a^\dag)\otimes \si_1+A^e.
\ee
Here $H_0$ is the Hamiltonian of the free field and a molecule without interaction,
$pV$ is the interaction Hamiltonian,
$\om_c>0$ is the cavity resonance frequency,  
$\om_a>0$ is the  molecular
frequency, and
  $p\in\R$ is proportional to the molecular dipole moment.
  The pumping is represented by a selfadjoint operator
   $A^e$, and
$\si_1$ and $\si_3$ are the Pauli matrices
%%%%%
 acting on the factor $\C^2$ in $F\otimes\C^2 $, so $[a,\si_k]=[a^\dag,\si_k]=0$.
%%%%%
Finally,
in (\ref{JC}),
$\ga>0$ and  $D$ is a dissipation operator.
We will consider the operator
\be\la{tiGa}
%D\rho&=& 
D_1\rho=a\rho a^\dag-\fr12 a^\dag a\rho-\fr12 \rho a^\dag a, 
 \ee
 used in \cite{A1973}--\cite{A1974}, \cite{BR1997,  BJ2007, L1976, SF2002,VW2006}, 
 and also its suitable modifications.
 %%%%%%%%%%%%%%%%%%%  
  \bd
  $\HS$ is the Hilbert space of Hermitian 
    Hilbert--Schmidt operators with the inner product {\rm \cite{RS1980}} 
\be\la{HS}
\langle\rho_1,\rho_2\rangle_\HS=\tr[\rho_1\rho_2].
\ee
\ed
%%%%%%%%%%%%%%%%%%%%%%%%%  
  \bd
i)  $|n,s_\pm\rangle=|n\rangle\otimes s_\pm$ form an orthonormal basis in $X$,
$s_\pm\in \C^2$ and $\si_3 s_\pm=\pm s_\pm$.
\smallskip\\
ii) $X_\infty$ is the space 
of finite linear combinations of the vectors $|n,s_\pm\rangle$.
\smallskip\\
iii) $\cD\subset\HS$ is the subspace of finite rank Hermitian operators 
\be\la{fr1}
\rho=\sum_{n,n'=0}^\infty \,\sum_{s,s'=s_\pm}\rho_{n,s;\,n',s'}
|n,s\rangle\otimes \langle n',s'|.
\ee
\ed

 Our main goal is to prove the well-posedness for the QRM
    in the  Hilbert space $\HS$  in the case of time-independent
pumping $A^e$.  The main issue is that
the operators $a$ and $a^\dag$ are unbounded by (\ref{aa}), so
the generator $\aA$ in QRM is also unbounded.
 Accordingly, the meaning of the QRM must be adjusted (see Definition \ref{gs}).

 The Hamilton operator $H$  is selfadjoint, so,
  in the case $\ga=0$, solutions are given by
  \be\la{rot}
  \rho(t)=e^{-i Ht}\rho(0)e^{iHt},\qquad t\ge 0.
      \ee
      In this case, the trace $\tr\!\rho(t)$ is conserved, and $\rho(t)\ge 0$ if $\rho(0)\ge 0$.
    However, for $\ga>0$ the formula  for solutions is missing.
 We    assume that \\ 
$\bullet$ The pumping $A^e$ and 
 the dissipation operator $D$
 are polynomials  (\ref{poldi}) in $a$ and $a^\dag$.\\
 $\bullet$ The dissipation operator $D$   and its adjoint $D^\dag$ are  nonpositive on 
 the  dense domain $\cD$:
    \be\la{np}
\langle\rho,D\rho\rangle_\HS\le 0,\qquad\langle\rho,D^\dag\rho\rangle_\HS\le 0,\qquad \rho\in\cD.
\ee
For example,
$A^e=a^\dag+a$, $A^e=a^\dag a$, and $D=D_1$  fulfill {the first  assumption}. 
As a key example, we prove in Theorem \ref{tm1} that
  the dissipation operator $D_1$ satisfies (\ref{np}).
 \smallskip

 Our main results are as follows. 
 \\
 I. The generator 
$\aA$ 
admits a closure $\ov\aA$ in $\HS$ from the domain $\cD$.
%and $\ov\aA$
%is
% {the}
This closure is
a generator of a strongly continuous contraction semigroup
in $\HS$.
\\
 II.  For all $\rho(0)\in\HS$, 
the trajectories $\rho(t)=e^{\ov\aA t}\rho(0)\in  C(0,\infty;\HS)$ are generalised solutions
to the QRM
% which are defined via matrix entries of $\rho(t)$
 (Definition \ref{gs}).
      \smallskip\\
    Our strategy is as follows.
   We  rewrite
 the QRM  as
\be\la{JC2}
\dot \rho(t)=\aA\rho(t):=K\rho(t)+\ga D\rho(t), 
\qquad{\rm where}\quad
K\rho=-i[H,\rho],\quad \rho\in\HS.
\ee
We   show that the operator
 $K$ is antisymmetric on  $\cD$:
    \be\la{Ksk}
\langle K\rho_1, \rho_2\rangle_\HS=-\langle\rho_1, K\rho_2\rangle_\HS, \qquad \rho_1,\,\rho_2\in \cD,
\ee
and that its quadratic form vanishes: 
\be\la{K02}
\langle\rho,K\rho\rangle_\HS=0,\qquad \rho\in \cD.
\ee
Then both operators $\aA$ and $\aA^\dag$
are nonpositive on $\cD$ by (\ref{np}). Hence,
 both operators $\aA$ and $\aA^\dag$ are dissipative, so
the existence of the semigroup follows from the Lumer--Phillips theorem \cite{LP1961}. The crucial
role in the proofs is played by
the 
polynomial structure (\ref{poldi}).

    \br\rm
The zero quadratic form (\ref{K02})
means that the vector field $K\rho$ is orthogonal to $\rho$
in the space $\HS$. Thus, the first term on the right hand side 
of (\ref{JC2}) can be interpreted as a rotations in $\HS$.
By (\ref{np}),
the second vector field $\ga D\rho$ is the generator of 
contractions
of $\HS$ which correspond to quantum spontaneous emission. 

\er

Let us comment on previous results in the field.
In the case of bounded generators $\aA$,
 semigroups for equations of type (\ref{JC}) obviously exist.
In this case,
Lindblad \cite{L1976} and  Gorini, Kossakowski, and Sudarshan \cite{GKS1976}
found necessary and sufficient conditions on $\aA$ providing the positivity and trace preservation.

For unbounded generators, the existence of dynamical semigroup 
for Quantum Dynamical Systems (QDS) is not well-developed, \ci[p.110]{AL2007}.
In \ci{D1977}, E.B. Davies considered quantum-mechanical Fokker--Planck 
equations (QFP).  
The existence of the corresponding positive 
contraction semigroup is established 
  in the Banach space of self-adjoint trace-class operators.
  The uniqueness and trace preservation have not been proved.
  Sufficient conditions, providing the trace preservation 
  for QFP equations,
  were 
  found in \ci{CF1998}. The detailed characterisation of a class of covariant 
  QDS with unbounded generators is presented in \ci{H1996}. 
  The survey
  can be found in \ci{D1976,F1999}.

  The theory \ci{GKS1976,L1976} 
  implies that for a class of dissipation operators,  
  (\ref{JC}) can be represented in the form of the QFP equation
  if $\aA$ is
  a bounded generator  of a completely positive semigroup (see Theorem 4.2 of \ci[Chapter 9]{D1976}).
  For unbounded generators, such a representation holds under
  appropriate  conditions \ci{EL1976}. 
 
We construct
 strongly continuous contraction 
 semigroups for  equation  (\ref{JC}) in the Hilbert space of the Hilbert--Schmidt
 operators
 which, in particular, contains all  trace class operators.
This framework allows us to employ the  well-developed theory of contraction semigroups
in the Hilbert space
by Lumer and Phillips and others.
 As a result, we prove that the semigroups exist under simple and easily verifiable
conditions
on the pumping and dissipation operators.
However,  the questions of positivity and trace preservation 
remain open;
 the needed development will be addressed elsewhere.

%%%%%%%%%%%%%%%%%%%%%%%%%%%%%%%%%%%%%%%%%%%%%%%%%%%%%
%%%%%%%%%%%%%%%%%%%%%%%%%%%%%%%%%%%%%%%%%%%%%%%%%%%%    
    \section{Notations and main results}
%%%%%%%%%%%%%%%%%%%%%%%%%%%%%%%%%%%%%%%%%%%%%%%%%%%%        

    A density operator $\rho\in\HS$
    is defined uniquely by its matrix entries
      \be\la{roent}
      \rho_{n,s;\,n',s'}=\langle n,s|\rho|n',s'\rangle, \qquad n,n'=0,1,\dots,\quad{s,s'=s_\pm}.
      \ee
     %which are Hermitian $2\times2$ matrices.
The Hilbert--Schmidt norm, corresponding to  the inner product  (\ref{HS}), can be written as
\be\la{HSen}
\Vert\rho \Vert_\HS^2=\tr[\rho^2]=\sum_{n,n'=0}^\infty
\sum_{s,s'=s_\pm}|\rho_{n,s;\,n',s'}|^2<\infty.
\ee
%Denote by $M_2=\C^2\otimes\C^2$ the space  of  
 %$2\times 2$-matrices.
Note that 
by (\ref{aa}), 
%\smallskip\\
%$\bullet$ 
the space
$X_\infty$ is invariant with respect to $a$ and $a^\dag$. Hence,
%\smallskip\\
%$\bullet$
the products 
of
$\rho\in\cD$ with any polynomials of $a$ and $a^\dag$ 
are well-defined as operators
in $X_\infty$.
%Hence, $D\rho\in\cD$ and  $\aA\rho\in\cD$ for $\rho\in\cD$ %by {\bf H2}.

Denote $M_2=\C^2\otimes\C^2$ the space  of 
 $2\times 2$-matrices.
 We assume that the pumping and the dissipation operator 
satisfy the following conditions.
\smallskip\\
{\bf H1.} $A^e$ and $D\rho$ are
polynomials in the creation and annihilation operators:
 \be\la{poldi}
 A^e=P(a,a^\dag),
\qquad 
D\rho=\sum_j Q_j\rho R_j,
 \ee
 where $P,Q_k,R_k$ 
 are polynomials in $a$ and $a^\dag$ with coefficients from $ M_2$. 
 \smallskip\\
{\bf  H2.}
  The operators $A^e$ 
and $D\rho$ with $\rho\in\cD$
 are  symmetric  in  $X_\infty$.
\smallskip\\
{\bf H3.} $D$ and $D^\dag$ are nonpositive in $\cD$
(that is, (\ref{np}) holds).

\br\rm
i) {\bf H1--H2} imply that
$\aA\rho\in\cD$ for 
$\rho\in\cD$.
\smallskip\\
ii) {\bf H3} holds if $Q_j=R_j^\dag$ for all $j$.

\er

%\bex \rm 
% {\bf H1--H2} hold for 
%$A^e=(a+a^\dag)\otimes\si_1$, $A^e=a^\dag a$,  $D=D_1$ and %$D=D_2$. {\bf H3--H4} hold  for $D=0$.

 %\eex
Our first result is as follows.

\bl\la{tm1} All conditions {\bf H1--H3} hold for the 
dissipation
operator $D=D_1$.

\el

To formulate other results, we need to
give a meaning to the QRM. The issue is that the operator $\aA$ is not well-defined on $\HS$.
Note that
the equation admits the treatment via the matrix 
entries (\ref{roent}) as the system
 \be\la{JCw} 
 \dot\rho_{n,s;\,n',s'}(t)=[\aA\rho(t)]_{n,s;\,n',s'},
 \qquad t\ge 0,\qquad n,n'\ge 0,\quad s,s'=s_\pm,
 \ee
since the right hand side is well-defined for all $\rho(t)\in\HS$.
Indeed,
by
{\bf H1} and (\ref{aa}),
 the generator $\aA$ is well-defined  on the domain $\cD$, and
 \be\la{JCwen} 
 [\aA\rho]_{n,s;\,n',s'}=
\sum_{\scriptsize\ba{c}|k-n|+|k'-n'|\le N\\ r,r'=s_\pm\ea} \aA_{n,s;\,n',s'}^{k,r;k',r'}\rho_{k,r;k',r'},\quad n,n'\ge 0, \quad s,s'=s_\pm,\quad\rho\in\cD,
 \ee
where $N=\max(2,\deg P, \max_j[\deg Q_j+\deg R_j])$.
Finally, since
the  summation in (\ref{JCwen}) is finite,
the matrix entries  $[\aA\rho]_{n,s;\,n',s'}$ admit a unique extension by continuity from 
$\rho\in\cD$
to all $\rho\in\HS$. Hence, 
the  operator $\aA$ admits a closed extension $\ov\aA$ from the domain $\cD$ to $\cD(\aA)\supset \cD$.
The structure (\ref{JCwen}) means that the matrix of the generator in the basis $|n,s\rangle$ is almost
diagonal.

\bd\la{gs}
We say that
a trajectory $\rho(t)\in 
C(0,\infty;\HS)$ is a generalised solution to the QRM if it 
satisfies the system 
(\ref{JCw}) in the sense of distributions; that is,
  \be\la{JCw2d} 
\rho_{n,s;\,n',s'}(t)-\rho_{n,s;\,n',s'}(0)
\!=\!
\int_0^t [\aA(\tau)\rho(\tau)]_{n,s;\,n',s'}
d\tau,\qquad t\ge 0, \quad\forall
\,\, n,n', s,s'.
 \ee
\ed

Our main  result is the following theorem.

\bt\la{tm2} Let  conditions {\bf H1--H3}  hold.
  Then 
    \smallskip\\
i)    $\ov\aA$ is a generator of a 
  strongly continuous contraction semigroup $U(t)=e^{\ov\aA t}$ in $\HS$,
and   for 
  \linebreak
  $\rho(0)\in\cD(\ov\aA)$, the trajectories  $\rho(t)=U(t)\rho(0)$ 
 are solutions to the equation
  \be\la{JCw2} 
\dot \rho(t)=
\ov\aA\rho(t),\qquad t\ge 0,
 \ee
 ii) For  all $\rho(0)\in \HS$, the  trajectories $\rho(t)=U(t)\rho(0)\in 
C(0,\infty;\HS)$
are generalised  solutions  to the QRM.

  \et

\setcounter{equation}{0}
\section{Nonpositivity of the dissipation operator and of its adjoint}
Here we prove Lemma \ref{tm1}.
Conditions {\bf H1} and {\bf H2} obviously hold. It remains to check
(\ref{np}).
First, let us calculate the adjoint operator 
 $D_1^\dag$: for  $\rho_1,\rho_2\in\cD$,
 \beqn\nonumber
\tr [\rho_1(D_1\rho_2)]
&=&\tr
\Big(
\rho_1(a\rho_2 a^\dag-\frac 12 a^{\dag}a\rho_2-\frac 12\rho_2 a^{\dag}a
%+a^{\dag}\rho_2a-\frac 12 aa^{\dag}\rho_2 -\frac 12 \rho_2  aa^{\dag}
)
\Big)
\\
\nonumber
&=&\tr\Big(
\rho_1a\rho_2 a^\dag-\frac 12 \rho_1a^{\dag}a\rho_2-\frac 12\rho_1\rho_2a^{\dag}a 
%+\rho_1a^{\dag}\rho_2a-\frac 12 \rho_1aa^{\dag}\rho_2 -\frac 12 \rho_1\rho_2  aa^{\dag}
\Big)\\
\nonumber
&=&\tr
\Big(
a^\dag\rho_1a\rho_2 -\frac 12 \rho_1a^{\dag}a\rho_2-\frac 12a^{\dag}a\rho_1\rho_2 %+a\rho_1a^{\dag}\rho_2-\frac 12 \rho_1aa^{\dag}\rho_2 -\frac 12  aa^{\dag}\rho_1\rho_2 
\Big)\\
\nonumber
&=&\tr
\Big((a^\dag\rho_1a -\frac 12 \rho_1a^{\dag}a-\frac 12a^{\dag}a\rho_1 
%+a\rho_1a^{\dag}-\frac 12 \rho_1aa^{\dag} -\frac 12  aa^{\dag}\rho_1
)\rho_2 
\Big)
=
\tr[(D_1^\dag\rho_1)\rho_2].
\eeqn
Hence, the adjoint operator $D_1^\dag$ differs from $D_1$
by swapping
$a$ and $a^\dag$:
\be\la{DD}
D_1^\dag\rho=a^\dag\rho a -\frac 12 \rho a^{\dag}a-\frac 12a^{\dag}a\rho,\qquad \rho\in\cD.
\ee
Second, let us prove 
the nonpositivity (\ref{np}) for $D=D_1$. For $\rho\in\cD$,
\beqn\la{dtr2}
\langle\rho,D_1\rho\rangle_\HS&=&\tr\big(\rho D_1\rho\big)
=\tr\Big(\rho\big(
a\rho a^\dag-\fr12 a^\dag a\rho-\fr12 \rho a^\dag a
%+a^\dag\rho a-\fr12 a a^\dag\rho-\fr12 \rho a a^\dag
\big)
\Big)
\nonumber\\
\nonumber\\
&=&
\tr\big(\rho a\rho a^\dag-\rho a^\dag a\rho
%+\rho a^\dag\rho a- \rho a a^\dag\rho
\big)
=\tr\big(\rho a\rho a^\dag- a^\dag a\rho^2
\big).
\eeqn
Now we use the fact that $\rho$ is a finite rank Hermitian operator (\ref{fr1}).
Then (\ref{poldi}) implies that
 the operators $\rho a\rho a^\dag$ and $a^\dag a\rho^2$
%and $a a^\dag\rho^2$ 
have only  finite 
number of nonzero entries (\ref{roent}), so their traces 
are well-defined. 
Moreover,
$\rho$ admits a finite spectral resolution  in the orthonormal basis of its eigenvectors  $e_i\in X_\infty$:
\be\la{eigro}
\rho=\sum_{i=1}^\nu\rho_i e_i\otimes e_i.
%|i\rangle\langle i|
\ee
In this basis, the entries  $\rho_{ij}=\rho_i\de_{ij}$, and 
the entries 
$a_{jk}=\langle e_j,a e_k\rangle$
and $a^\dag_{kl}=\langle e_k,a^\dag e_l\rangle$
of the operators $a$ and $a^\dag$ are well-defined.
Hence, (\ref{dtr2}) implies,
with summation in repeated indices, 
\beqn\la{dtr3}
\langle\rho,D_1\rho\rangle_\HS&=&
\rho_i\de_{ij}a_{jk}\rho_k\de_{kl}
a^\dag_{li}
%\rho_l(t)\de_{li}
-a^\dag_{kl}a_{lj}\rho_j^2\de_{jk}
%-a_{jk}a^\dag_{kl}\rho_l^2\de_{lj}
%
=
\rho_i a_{ik}\rho_k
a^\dag_{ki}
%???
-a^\dag_{kl}a_{lk}\rho_k^2
%-a_{jk}a^\dag_{kj}\rho_j^2
\nonumber\\[1ex]
&=&
\rho_i a_{ik}\rho_k
a^\dag_{ki}
%???????????
-a^\dag_{ki}a_{ik}\rho_k^2
%-a_{ik}a^\dag_{ki}\rho_i^2
=
a_{ik}a^\dag_{ki}(\rho_i\rho_k
-\rho_k^2
%-\rho_i^2
)
\nonumber\\
&=&
\fr12\Big(a_{ik}a^\dag_{ki}(\rho_i\rho_k
-\rho_k^2)+a_{ki}a^\dag_{ik}(\rho_k\rho_i
-\rho_i^2) 
)
=
-\fr12
|a_{ik}|^2(\rho_i-\rho_k)^2
\le 0
\eeqn
since $a^\dag_{ik}=\ov a_{ki}$. Hence,  the nonpositivity  is proved for $D_1$.
For $D_1^\dag$ the proof is the same with the swapping
$a$ and $a^\dag$.

\br\rm
The proof of the nonpositivity 
   essentially depends on the symmetry of  $\rho$.
 \er

\setcounter{equation}{0}
\section{Semigroup and generalised solutions}

\bl\la{lnp}
Both operators $\aA$ and $\aA^\dag$  are well-defined and nonpositive on the domain $\cD$.

\el
\bpr i)
 For $\rho_1,\rho_2\in\cD$, we have in the notation (\ref{JC2}):
\beqn\la{fr2}
\langle\rho_1, K\rho_2\rangle_\HS
=-i\tr(\rho_1[H,\rho_2])=-i\tr(\rho_1(H\rho_2-\rho_2H))
%\nonumber\\
%\nonumber\\
=
-i\tr(\rho_1H\rho_2-\rho_2H\rho_1).
%={\cre ??????????}\langle K\rho_1, \rho_2\rangle_\HS.
\eeqn
where all the terms are well-defined by {\bf H1}.
Hence, (\ref{K02}) holds.
%$K$ is a skewsymmeric operator on $\cD$, and
%\be\la{K0}
%\langle\rho, K\rho\rangle_\HS= 0, \qquad \rho\in \cD.
%\ee
Therefore, $\aA=K+\ga D$ is nonpositive by {\bf H3}.
Similarly, we obtain (\ref{Ksk}):
\beqn\la{KK}
\langle K\rho_1, \rho_2\rangle_\HS
&=&i\tr([H,\rho_1]^\dag\rho_2)=-i\tr((H\rho_1-\rho_1H)\rho_2)
=-i\tr(\rho_1\rho_2H-\rho_1H\rho_2)
\nonumber\\
\nonumber\\
&=&
i\tr(\rho_1[H,\rho_2])
=-\tr(\rho_1 (K\rho_2))=-\langle \rho_1, K\rho_2\rangle_\HS.
\eeqn
Hence  $\aA^\dag|_\cD=-K+\ga D^\dag$, so  $\aA^\dag$ is also
well-defined on $\cD$ and
 nonpositive by {\bf H3}. 
\epr

\noindent
{\bf Proof of Theorem \ref{tm2}.}
{\it i)}
%%AC what is this "ad" ????
Lemma \ref{lnp} together with Proposition \ci[II.3.23]{EN2000} imply 
that both operators $\aA$ and $\aA^\dag$ are dissipative. Moreover, 
the  operator $\aA$ is densely defined and admits a closed extension $\ov\aA$ by (\ref{JCwen}).
Now Theorem \ref{tm2} i) follows
from  \ci[Corollary II.3.17]{EN2000}
 which is a corollary of 
the 
Lumer--Phillips theorem
\ci[II.3.15]{LP1961}. 
\smallskip\\
{\it ii)} 
The contraction means that for $\rho(0)\in\HS$, we have:
  \be\la{HS3}
\Vert\rho(t)\Vert_\HS\le \Vert\rho(0)\Vert_{\HS},\qquad t\ge 0.
\ee
To prove the identity (\ref{JCw2d}), let us take 
$\rho_k(0)\in\cD(\ov\aA)$ and
$\rho_k(0)\toHS\rho(0)$. Then 
also
$\rho_k(t)\in\cD(\ov\aA)$ and 
$\rho_k(t)\toHS\rho(t)$
uniformly for  $t\ge 0$ by (\ref{HS3}). 
The equation (\ref{JCw2}) for
 $\rho_k(t)$,
implies the corresponding equations (\ref{JCw}) since
the formulas (\ref{JCw}) hold for 
$\rho\in\cD(\ov\aA)$.
Hence, (\ref{JCw2d}) for $\rho(t)$ follows
from similar  integral identities for $\rho_k(t)$
as $k\to\infty$
 since the summations are finite for every $n,s,n',s'$.

\section{Acknowledgements} 
The authors thank S. Kuksin, M.I. Petelin, A. Shnirelman, and H. Spohn
 for long-term fruitful discussions, and the  Institute of Mathematics of BOKU University
for the support and hospitality. 
The research is supported by
Austrian Science Fund (FWF) PAT 3476224.

\section{Conflict of interest}
We have no conflict of interest.

 \section{Data availability statement}
The manuscript has no associated data.

%\end{document}

\end{document}